\documentclass[conference]{IEEEtran}
\IEEEoverridecommandlockouts
\usepackage{cite}
\usepackage{amsmath,amssymb,amsfonts}
\usepackage{algorithmic}
\usepackage{graphicx}
\usepackage{textcomp}
\usepackage{xcolor}
\usepackage{siunitx}
\usepackage{comment}

\usepackage[normalem]{ulem}

\def\BibTeX{{\rm B\kern-.05em{\sc i\kern-.025em b}\kern-.08em
    T\kern-.1667em\lower.7ex\hbox{E}\kern-.125emX}}
\begin{document}

\title{Few-Shot Adaptation for Anomalous Sound Detection at Low False-Positive Rate}

\author{
\IEEEauthorblockN{Simone Lugani\IEEEauthorrefmark{1}, Christof Pichler\IEEEauthorrefmark{2}, Hannes Wegleiter\IEEEauthorrefmark{2}, Christoph Feilmayr\IEEEauthorrefmark{3}, Rodolfo Zunino\IEEEauthorrefmark{1}, Edoardo Ragusa\IEEEauthorrefmark{1}}
\IEEEauthorblockA{\IEEEauthorrefmark{1}DITEN, University of Genoa, Genoa, Italy; contact: simone.lugani@edu.unige.it}
\IEEEauthorblockA{\IEEEauthorrefmark{2}Christian Doppler Laboratory for Measurement Systems for Harsh Operating Conditions, EMS, TU Graz, Graz, Austria}
\IEEEauthorblockA{\IEEEauthorrefmark{3} voestalpine Stahl GmbH, Linz, Austria}
}

\IEEEpubid{%
  \parbox[t]{\columnwidth}{%
    \footnotesize
    \raggedright
    \copyright~2026 IEEE.
    Personal use of this material is permitted.
    Permission from IEEE must be obtained for all other uses, in any current
    or future media, including reprinting/republishing this material for
    advertising or promotional purposes, creating new collective works, for
    resale or redistribution to servers or lists, or reuse of any copyrighted
    component of this work in other works.\\[1pt]
    \emph{Published in IEEE COINS 2026.\\}
    DOI: 10.1109/COINS69995.2026.11715629.
  }%
  \hspace{\columnsep}%
  \makebox[\columnwidth]{}%
}
\maketitle

\begin{abstract}
Reliable fault detection is crucial for industrial machinery in large-scale facilities, where unexpected failures can cause costly downtime and production interruptions. Although acoustic monitoring provides a scalable solution for machine condition monitoring, its deployment is often constrained by high false-positive rates. We propose a two-step approach to automatic anomalous sound detection: first, when only normal operating audio samples are available, a regularized unsupervised model is deployed; then, after the first machine fault, a few-shot adaptation strategy exploits the anomalous samples acquired during the fault to improve the system. This adaptation increases the area under the precision-recall curve (AUPR) from 23.0 to 64.0 and improves detection in the low false-positive-rate region, with the partial area under the receiver operating characteristic curve (pAUROC) increasing from 39.6 to 72.8. We also show that covariance regularization improves both prediction accuracy and training stability in the unsupervised model. These results provide a practical strategy for acoustic anomaly detection systems, enabling improved accuracy and reduced false alarms.
\end{abstract}

\renewcommand\IEEEkeywordsname{Keywords}


\section{Introduction}

Reliable operation of industrial machinery is essential in continuous-process industries, where unexpected failures can cause costly downtime and production interruptions~\cite{Yamamoto2016}. This is vital in pig iron production, where the blast furnace stock house prepares raw materials before charging, and large vibrating screens make early fault detection essential for reliable operation. While vibration-based monitoring is well established, its use in large-scale facilities is often limited by the required sensor effort and installation cost~\cite{Hashemian2011B}. Acoustic condition monitoring uses microphones to capture machine-related sound emissions~\cite{Brooks2018}. In stock house environments, however, multiple concurrently operating screens, time-varying background noise, and additional disturbances can mask weak fault-related signatures, making reliable detection difficult~\cite{jombo2023acoustic}.

Instead of requiring large labelled fault datasets, unsupervised anomalous sound detection (ASD) systems learn normal machine behaviour and identify deviations as anomalies~\cite{jombo2023acoustic}. In recent years, this field has advanced strongly, in particular through the DCASE challenge series, where a wide range of data-driven methods for machine condition monitoring has been developed~\cite{di2022anomalous,suefusa2020anomalous,jalali2022dcase,wang2023abnormal,zeng2022robust,chinnasamy2024acoustic}. Recent DCASE tasks have also moved towards first-shot and domain-generalisation settings, which are highly relevant for practical applications~\cite{nishida2025description,wilkinghoff2025handling,wilkinghoff2025local,saengthong2025deep,yan2024transformer}. However, a gap remains between current ASD research and the requirements of real industrial applications. In large-scale facilities, the system must operate at very low false-positive rates (FPR), since even a small number of false alarms can limit the usability and acceptance of monitoring systems. At the same time, real-world acoustic conditions are highly complex due to multiple simultaneously operating machines, changing operating states, and additional disturbing sound sources~\cite{jombo2023acoustic}. While ASD has already been applied to such harsh acoustic environments~\cite{U_ASD_2025,pichler2023acoustic}, the resulting FPRs often remain a limiting factor for reliable deployment. In addition, a small number of fault samples is often available after the first observed failure, creating an opportunity to improve the separation between normal and faulty conditions and, in particular, to reduce false alarms. This setting can be viewed as a few-shot adaptation scenario, where only limited anomalous data are available for model adaptation. 

\begin{figure}[t]
\centering
\includegraphics[width=0.85\linewidth]{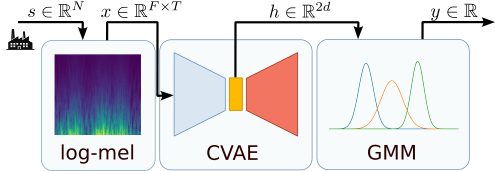}
\caption{Block diagram of the ASD framework: the log-mel spectrogram is compressed by the CVAE; then, a GMM is used to compute the anomaly score based on vector similarity with the learned normal representation.}
\label{fig:ASD_framework}
\vspace{-0.5cm}
\end{figure}

We propose a two-stage ASD framework for industrial applications that enables targeted refinement of an unsupervised model using a small number of fault samples. First, a convolutional variational autoencoder (CVAE) extracts the features, and a Gaussian mixture model (GMM) models the normal operating condition using only normal data. Later, after the first observed fault, the CVAE is fine-tuned with a few anomalous samples from a single machine fault, while the GMM is kept fixed as a reference for the normal latent distribution. This refinement improves the separation between normal and faulty conditions, with focus on the low-FPR operating region, after the first machine fault. In addition, we investigate covariance regularization in the GMM, which improves robustness in the unsupervised setting. The evaluation is conducted on real-world acoustic data from an operating industrial blast furnace stock house, demonstrating relevance under realistic deployment conditions. The proposal outperforms the baselines in terms of overall AUPR and detection capabilities under stringent false-alarm requirements.

\section{Methods}

We propose a two-stage learning paradigm for ASD in industrial facilities. Section~\ref{sec:ASD_Framework} describes the ASD framework. Initially, when only normal operating data are available, an unsupervised anomaly detection model is trained and deployed, as described in Section~\ref{sec:unsupervised}. After the first machine fault occurs, the recorded sound data is used to enhance the model through a few-shot adaptation step integrating the newly acquired knowledge, as described in Section~\ref{sec:one-shot}.

\subsection{Anomalous Sound Detection Framework}
\label{sec:ASD_Framework}
Figure~\ref{fig:ASD_framework} illustrates the ASD framework. Audio samples from the facility are represented by their log-mel spectrograms. An undercomplete autoencoder, specifically a CVAE, compresses normal audio signals into a lower-dimensional latent space, which is then modeled by a GMM. During inference, the anomaly score of a new audio sample is defined as the negative log-likelihood of its latent representation under the GMM, so that higher scores indicate anomalous behaviour.

\subsubsection{Preprocessing}
For each input audio segment $s \in \mathbb{R}^{N}$, the log-mel spectrogram $x \in \mathbb{R}^{F \times T}$ is computed and normalized to the [0,1] range using Min-Max normalization.

\subsubsection{Convolutional Variational Autoencoder}
The CVAE receives as input the log-mel spectrogram $x$ extracted in the preprocessing stage and learns a compact representation of normal audio patterns in the latent space with dimensionality $d$. The encoder outputs two parameters for each latent dimension, namely the mean and the log-variance. Therefore, the latent representation is defined by a mean vector $\mu \in \mathbb{R}^{d}$ and a log-variance vector $\log \sigma^{2} \in \mathbb{R}^{d}$.

\subsubsection{Gaussian Mixture Model Anomaly Scoring}
The normal latent representation extracted by the CVAE is modeled by a GMM. For each input sample, the mean and log-variance vectors estimated by the CVAE are concatenated into the feature vector $h = [\mu; \log \sigma^2] \in \mathbb{R}^{2d}$, which is used as input to the GMM. During inference, the anomaly score $y \in \mathbb{R}$ is computed as the negative log-likelihood of $h$ under the GMM, so that higher values correspond to a higher degree of anomaly.

\subsection{Unsupervised Training}
\label{sec:unsupervised}
First, an unsupervised model is trained with only normal operating data. The CVAE is trained to compress and reconstruct log-mel spectrograms of normal audio samples using the standard variational objective: a reconstruction term encourages accurate reconstruction, and a Kullback--Leibler (KL) regularization term promotes a smoother and more stable latent space. The unsupervised CVAE loss is defined as:

\begin{equation}
\mathcal{L}_{\mathrm{unsup}} =
\|x-\hat{x}\|^2
- \frac{\beta}{2}\sum_{i=1}^{d}
\left(1+\log\sigma_i^2-\mu_i^2-\sigma_i^2\right)
\end{equation}

\(x\) and \(\hat{x}\) denote the input spectrogram and its reconstruction, \(\mu_i\) and \(\sigma_i^2\) denote the mean and variance of the \(i\)-th latent dimension, and $\beta$ is the weight of the KL regularization term.

Once the CVAE is trained, the GMM is fitted to the latent representations extracted from a dedicated set of normal recordings. For each sample, the mean and log-variance vectors are concatenated to form the GMM input features. Unlike previous works~\cite{U_ASD_2025}, covariance regularization is used during GMM training to obtain a more robust estimate of the latent data distribution. The resulting unsupervised model, combining the CVAE encoder with the GMM in the latent space, defines the first stage of the ASD system and serves as the starting point for the subsequent learning phase.

\subsection{Few-shot Adaptation with Single Machine Fault}
\label{sec:one-shot}

The second training stage improves the ASD model after the first observed fault. The unsupervised CVAE is fine-tuned using a small set of anomalous samples collected from a single fault event. The objective is not to specialize the model to a specific machine, but to use real anomaly samples to improve the separation of normal and anomalous sounds in the facility.

The GMM obtained during the unsupervised stage is kept as the reference model for the normal latent distribution. This preserves a stable estimate of the normal latent distribution and prevents drift during fine-tuning. Each anomalous sample is used to generate a small number of augmented samples through minor temporal shifts, temporal cropping, amplitude scaling, and spectral coloration. The CVAE is then optimized using both normal and augmented anomalous samples. The loss function separates normal and anomalous samples in the latent space, while preserving the reconstruction behavior on normal data as a regularization term. The overall loss for the few-shot adaptation stage is defined as:

\begin{equation}
\mathcal{L}_{\mathrm{FS}} = \mathcal{L}_{\mathrm{Z,\, N}} + \mathcal{L}_{\mathrm{Z,\, A}} + \lambda \mathcal{L}_{\mathrm{unsup}}
\end{equation}

\( \mathcal{L}_{\mathrm{Z,\, N}} \) enforces consistency of normal samples with the baseline latent distribution, \( \mathcal{L}_{\mathrm{Z,\, A}} \) promotes the separation of anomalous samples from the GMM-based representation of normal data in the latent space, and \( \mathcal{L}_{\mathrm{unsup}} \) denotes the standard VAE loss as defined in Sec. \ref{sec:unsupervised}. \( \mathcal{L}_{\mathrm{Z,\, N}} \) and \( \mathcal{L}_{\mathrm{unsup}} \) are computed for normal data only, and \( \mathcal{L}_{\mathrm{Z,\, A}} \) for anomalous data only. \( \lambda \) controls the contribution of the unsupervised term.

Specifically, the two latent-space terms are defined as:

\begin{equation}
\mathcal{L}_{\mathrm{Z,\, N}} =
\max\left(0, \operatorname{GMM}(h(x)) - m_{\mathrm{N}}\right)^2
\end{equation}
\begin{equation}
\mathcal{L}_{\mathrm{Z,\, A}} =
\max\left(0, m_{\mathrm{A}} - \operatorname{GMM}(h(x))\right)^2
\end{equation}

where \( m_{\mathrm{N}}\) and \( m_{\mathrm{A}}\) denote the margins for normal and anomalous samples, respectively. They were selected manually from the pre-adaptation distribution of normal anomaly scores, without optimization on anomalous validation data, as no independent fault-validation set was available and using test faults would introduce data leakage.

\section{Experiments}
\subsection{Data Acquisition and Datasets}

An acoustic measurement system was deployed in an operating blast furnace stock house to collect a custom real-world dataset under realistic industrial conditions~\cite{pichler2023acoustic}. Recordings were acquired over several months at \SI{48}{\kilo\hertz} using a Behringer ECM8000 condenser microphone connected to a Zoom F8n MultiTrack Field Audio Recorder and positioned approximately \SI{5}{\meter} from the vibrating screen line. The acoustic environment is characterized by high sound pressure levels of approximately \SI{95}{\decibel_{SPL}} and multiple concurrent sound sources, including \num{13} vibrating screens under varying load conditions and additional non-stationary disturbances. All recordings include realistic background disturbances and were intentionally used without preprocessing to evaluate the system under practical deployment conditions. The normal operating dataset comprises \SI{60000}{\second} of data for CVAE training, \SI{60000}{\second} for GMM fitting, and \SI{71200}{\second} of independent normal test data.

\begin{figure}[t]
\centering
\includegraphics[width=1\linewidth]{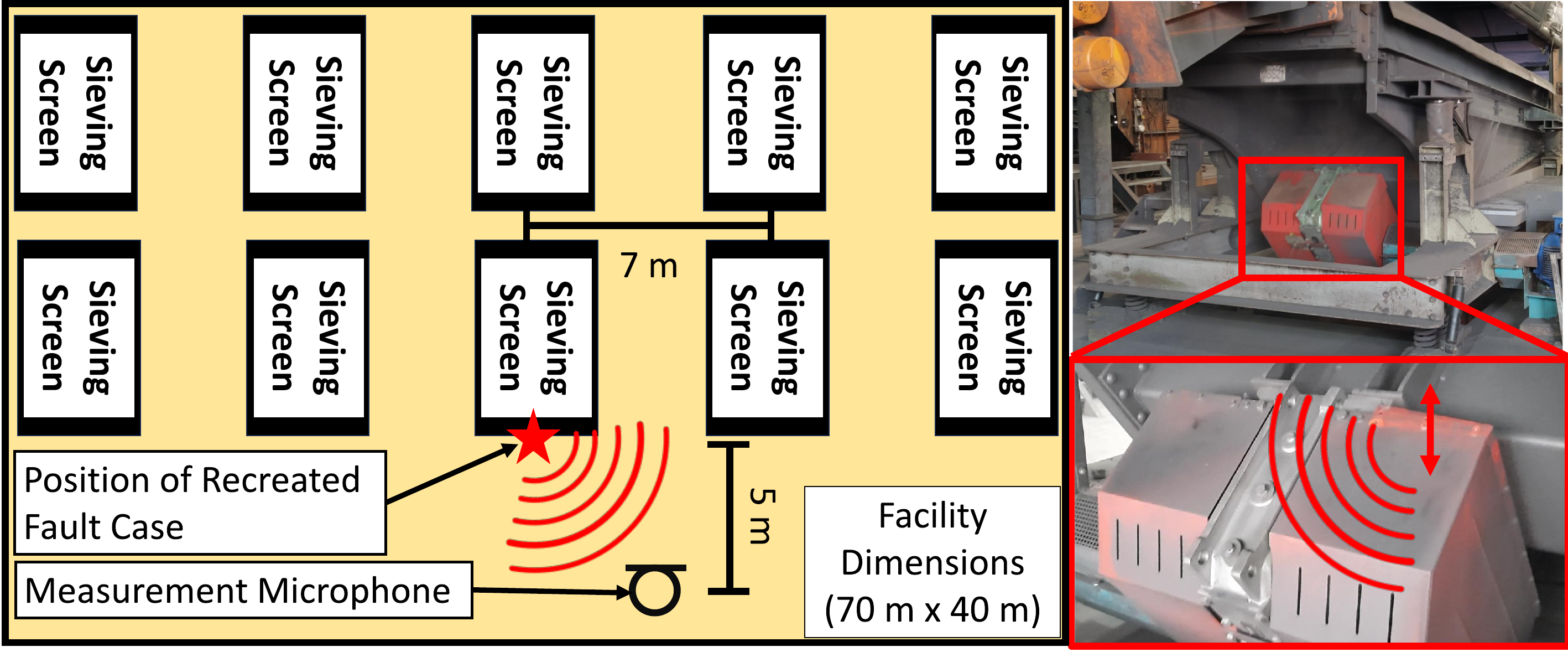}
\caption{Overview of the measurement setup and recreated fault case: facility layout with microphone and fault locations, vibrating screening machine, and physical reconstruction of typical mechanical failures.}
\label{fig:facility}
\vspace{-0.5cm}
\end{figure}

Since real fault events are rarely available in sufficient quantity, dedicated on-site measurement campaigns were conducted with experienced plant personnel to physically recreate representative fault cases at multiple locations within the facility, including ruptured springs and detached metal plates causing characteristic metallic impact sounds. Figure~\ref{fig:facility} overviews the acquisition setup, and an example recreated fault case in which a detached metal plate emulates a typical mechanical failure scenario. The resulting real-world fault dataset comprises \SI{400}{\second} of recordings, acquired with one of three possible faulty machines running. These samples were divided into folds to train the models on a few samples from a single machine and test the method on the other two, as described in Sec. \ref{sec:experimental_setup}.

\subsection{Experimental setup}
\label{sec:experimental_setup}

Input audio recordings were segmented into \SI{1.37}{\second} samples with a hop size of \SI{1.0}{\second}. Log-mel spectrograms used an FFT length of 1024, hop size of 512, and 64 mel filter banks between 500 and 10\,000 Hz, resulting in \(64 \times 128\) inputs to the CVAE. The CVAE used a 16-dimensional latent space with five-layer convolutional encoder and transposed-convolutional decoder. We fixed \(\beta=0.001\), \(\lambda=0.1\), \(m_\mathrm{N}=-20\), and \(m_\mathrm{A}=500\). The CVAE was trained with Adam, learning rate \(10^{-3}\) and early stopping. The GMM was fitted on a separate set of normal latent representations using full covariance matrices, 50 components, covariance regularization \(10^{-5}\), and 10 random initializations. During few-shot adaptation, the CVAE was initialized from the unsupervised model and fine-tuned with Adam using learning rate \(10^{-5}\), early stopping, and exponential learning-rate decay with factor 0.95; fine-tuning batches were balanced between normal and anomalous samples, and batch-normalization layers were frozen. All models were implemented in PyTorch and trained on an NVIDIA RTX 4080 GPU. To account for result variability, each model was trained 10 times with different random seeds, and all metrics are reported as mean and standard deviation.

Normal data were divided into three sets for CVAE training, GMM fitting, and model testing. The anomalous data were collected from three faulty machines and divided into corresponding folds according to the machine where the fault was introduced. The recordings associated with the three machines amounted to \SI{120}{\second}, \SI{220}{\second}, and \SI{60}{\second}, respectively, and only one machine was used for few-shot adaptation in each configuration. The experiments were repeated three times, each time using two folds as the test set and the remaining one for few-shot training. Accordingly, \(F_x\) denotes the configuration where fold \(x\) is excluded from the test set and used to train the few-shot model. Global metrics are reported as the average over the three fold configurations.

\begin{figure}[t]
\centering
\includegraphics[width=0.95\linewidth]{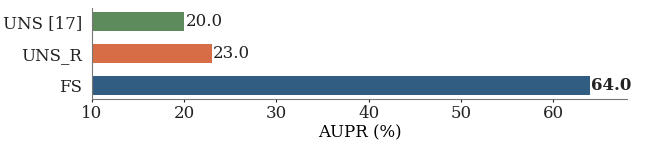}
\caption{AUPR performance of the ASD model with regularization after the unsupervised training step (UNS\_R) and after few-shot adaptation (FS), compared with the unsupervised state-of-the-art approach without regularization (UNS). The few-shot adaptation step yields a drastic improvement.}
\label{fig:sota}
\end{figure}

\subsection{Comparison with State-of-the-Art}

Figure~\ref{fig:sota} shows the area under the precision--recall curve (AUPR) obtained by three methods. UNS denotes the unsupervised baseline from~\cite{U_ASD_2025}, which combines an autoencoder and a GMM for unsupervised ASD; unlike the original work, it is implemented here with a CVAE instead of a fully connected architecture, consistently with the framework in Sec.~\ref{sec:ASD_Framework}. UNS\_R uses the same approach with covariance regularization in the GMM, while FS denotes the proposed framework after few-shot adaptation. Comparing UNS and UNS\_R shows that covariance regularization improves the average AUPR from 20.0 to 23.0 and reduces the standard deviation over 10 training runs, not explicitly reported in the figure, from \(8.0\) to \(6.2\), indicating both higher mean performance and more stable training. The subsequent few-shot adaptation further increases the AUPR from \(23.0\) to \(64.0\), showing that the proposed adaptation step effectively exploits the limited fault information available after the first observed anomaly.

\subsection{Performance at Low False-Positive-Rate}
\begin{table}[t]
    \centering
    \caption{Performance comparison of the models (mean $\pm$ std)}
    \label{tab:low_fpr_model_comparison}
    \renewcommand{\arraystretch}{1.4}
    \setlength{\tabcolsep}{8pt}
    \setlength{\arrayrulewidth}{0.6pt}
    \begin{tabular}{|c|c|c|c|}
        \hline
        \textbf{Model} & \textbf{pAUROC} & \textbf{TPR@FPR 1\%} & \textbf{TPR@FPR 0.1\%} \\
        \hline
        UNS & $35.5 \pm 1.2$ & $58.0 \pm 1.4$ & $7.5 \pm 0.8$ \\
        \hline
        UNS\_R & \uline{$39.6 \pm 0.9$} & \uline{$64.4 \pm 1.0$} & \uline{$9.4 \pm 0.8$} \\
        \hline
        FS & $\mathbf{72.8 \pm 0.4}$ & $\mathbf{89.3 \pm 0.0}$ & $\mathbf{60.6 \pm 0.1}$ \\
        \hline
    \end{tabular}
\end{table}

Table~\ref{tab:low_fpr_model_comparison} reports low-FPR performance as mean and standard deviation for each ASD model. The pAUROC is computed over the region with FPR \(\leq 1\%\), while TPR@FPR 1\% and TPR@FPR 0.1\% denote the true positive rate obtained by fixing the threshold at FPR \(=1\%\) and \(=0.1\%\), respectively.

The FS model shows that the proposed adaptation step substantially improves low-FPR performance: the system becomes more reliable at FPR \(=1\%\) and can operate at FPR \(=0.1\%\), detecting more than 1 out of 2 anomalous samples. This makes the model more suitable for very low-FPR operation, especially when sample-level predictions are combined through binary integration~\cite{richards2005fundamentals}.
Across all metrics, UNS\_R provides a small but consistent improvement in mean value and variability over the baseline UNS. However, its TPR is limited at FPR \(=1\%\) and remains unreliable at FPR \(=0.1\%\), detecting on average fewer than 1 out of 10 anomalous samples.

\begin{table}[t]
    \centering
    \caption{pAUROC for each fold (mean $\pm$ std)}
    \label{tab:low_fpr_model_comparison_per_fold}
    \renewcommand{\arraystretch}{1.4}
    \setlength{\tabcolsep}{8pt}
    \setlength{\arrayrulewidth}{0.6pt}
    \begin{tabular}{|c|c|c|c|}
        \hline
        \textbf{Model} & \textbf{$F_1$} & \textbf{$F_2$} & \textbf{$F_3$} \\
        \hline
        UNS & $42.4 \pm 12.3$ & $19.6 \pm 9.4$ & $44.4 \pm 14.0$ \\
        \hline
        UNS\_R & \uline{$45.9 \pm 8.2$} & \uline{$23.3 \pm 8.5$} & \uline{$49.7 \pm 10.2$} \\
        \hline
        FS & $\mathbf{75.2 \pm 2.6}$ & $\mathbf{64.5 \pm 0.5}$ & $\mathbf{78.7 \pm 8.8}$ \\
        \hline
    \end{tabular}
\end{table}

Table \ref{tab:low_fpr_model_comparison_per_fold} reports the pAUROC results by fold, with one column for each fold. The metrics under $F_{1}$ are obtained by adapting the model using only anomalies from the first machine and testing the ASD performance on the second and third faulty machines. The same applies to $F_{2}$, where the second machine is used for training, and to $F_{3}$, where the third is used, always testing on the other two. The previously discussed results are confirmed and consistent, indicating that this approach can exploit limited anomalous samples from one machine to improve detection on the remaining machines in the considered facility. Nevertheless, this table shows how the train–test data split can affect absolute values, especially when the number of samples is very small, as in this case.

\section{Conclusion}
Covariance regularization improves the unsupervised GMM baseline in both average performance and stability. The proposed few-shot adaptation step leads to a substantial gain in detection capability, especially in the low FPR operating region, which is the most relevant for real industrial deployment, yielding higher pAUROC and TPR at very low FPR values, making the system much more effective in scenarios where false alarms must be strictly limited.

Overall, the proposed adaptation paradigm is promising for industrial ASD, as it allows the model to exploit newly observed faults and improve future detection. Future work will extend the evaluation to additional fault types, longer deployments, and hyperparameter sensitivity analysis.

\section*{Acknowledgment}
This work was supported in part by the Austrian Federal Ministry of Digital and Economic Affairs, in part by the National Foundation for Research, Technology and Development, Austria, in part by the Christian Doppler Research Association, and in part by TU Graz Open Access Publishing Fund, Austria.

\bibliographystyle{IEEEtran}
\bibliography{references}
\vfill
\end{document}